\documentclass[prb,referee,floatfix,superscriptaddress,amsmath,amssymb,twocolumn,showpacs]{revtex4-1}

\usepackage{pstricks}
\usepackage[latin1]{inputenc}
\usepackage{graphics}
\usepackage{graphicx}
\usepackage{amssymb}
\usepackage{amsfonts}
\usepackage{amsmath}
\usepackage{color}
\usepackage{bm}
\usepackage{epsf}

\def \ba {\begin{eqnarray}}
\def \ea {\end{eqnarray}}

\newcommand{\op}[1]{\mathbf{#1}}
\newcommand{\ve}[1]{\vec{#1}}

\def\prl#1#2#3{Phys.\ Rev.\ Lett.\ {\bf #1}, #2 (#3)}

\def\prb#1#2#3{Phys.\ Rev.\ B {\bf #1}, #2 (#3)}

\def\jpa#1#2#3{J.\ Phys.\ A {\bf #1}, #2 (#3)}

\def\jpsj#1#2#3{J.\ Phys.\ Soc.\ Jpn.\ {\bf #1}, #1 (#3)}

\begin{document}

\title{Combined analytical and numerical approach to study magnetization 
 plateaux in doped quasi-one-dimensional antiferromagnets}

\author{C.A.\ Lamas}
\affiliation{Laboratoire de Physique Th\'eorique,  IRSAMC, CNRS and Universit\'e de Toulouse, UPS, F-31062 Toulouse, France}

\author{S. Capponi}
\affiliation{Laboratoire de Physique Th\'eorique,  IRSAMC, CNRS and Universit\'e de Toulouse, UPS, F-31062 Toulouse, France}

\author{P.\ Pujol}
\affiliation{Laboratoire de Physique Th\'eorique,  IRSAMC, CNRS and Universit\'e de Toulouse, UPS, F-31062 Toulouse, France}

\date{\today}


\begin{abstract}
We investigate the magnetic properties of  quasi one-dimensional 
quantum spin-$S$ antiferromagnets. We use a combination 
of analytical and numerical techniques to study the presence 
of plateaux in the magnetization curve. The analytical technique 
consists in a path integral formulation in terms of coherent states. 
This technique can be extended to the presence of doping and 
has the advantage of a much better control for large spins than 
the usual bosonization technique. We discuss the appearance 
of doping-dependent plateaux in the magnetization curves 
for spin-$S$ chains and ladders. The analytical results are 
complemented by a Density matrix Renormalization Group (DMRG) 
study for a trimerized spin-$1/2$ and anisotropic spin-$3/2$ 
doped chains. 
\end{abstract}

\maketitle

\section{Introduction}
\label{sec:intro}

One-dimensional and quasi one-dimensional  quantum spin systems 
have been a subject of close attention in condensed matter physics. 
Over the past decades, much progress has emerged from the studies 
of such low-dimensional  systems, the Haldane gap \cite{Haldane} 
and the magnetization plateaux \cite{OYA, plateaux_ladd, Plateaux_pmer} 
being some of the best-known examples. In particular, spin chains 
have allowed to study quantum magnetism in the simplest tractable setting,  
while investigations of spin ladders have permitted first steps 
in the study of crossover from one to two dimensions.

The present work is devoted to magnetization plateaux in quantum 
antiferromagnets: the magnetization remaining spectacularly constant 
in a finite interval of external magnetic field. This phenomenon has 
been found in a great variety of systems from spin ladders 
\cite{plateaux_ladd} and $p$-merized chains \cite{Plateaux_pmer} 
to frustrated higher dimensional systems. \cite{penc}.

For a quantum spin-$S$ chain, the necessary  condition for the occurrence 
of a magnetization plateau has been established by Oshikawa, Yamanaka 
and Affleck \cite{OYA}:
\ba
\label{eq:condition_intro_1}
N(S - m) \in \mathbb{Z}, 
\ea
where $m$ is the  magnetization per site, $N$ is the number of spins 
per unit cell, and $\mathbb{Z}$ is the set of all integers.

This condition restricts the plateau magnetization $m$ to rational values. 
However, it has been argued that plateaux may also appear at an irrational $m$, 
as a result of either quenched disorder \cite{Plateau_disorder} or doping 
with itinerant carriers 
\cite{Cabra_irrational,Poilblanc2004,Roux_irrational,Roux_2007,Roux_PhD}. 
In the latter case, doping may allow to reduce the $m$ in a 
controlled way, thus making the plateau more easily accessible 
to experiments in lower magnetic fields.

Unfortunately, the analytical methods, used so far to study magnetization 
plateaux in spin chains and ladders, have been limited to bosonization 
(and thus, effectively, to one-dimensional spin-$1/2$ systems) -- and 
to the bond operator technique, intrinsically restricted to spin-$1/2$ systems \cite{MomTot}.  
However, recently, Tanaka, Totsuka and Hu \cite{Tanaka} (TTH) have arrived 
at the necessary condition of Ref.~\onlinecite{OYA} using the Haldane's 
path integral approach. This approach is applicable regardless of the 
dimensionality or of the value $S$ of the spin -- and, below, we explore 
how much progress it may afford us in understanding the magnetization 
plateaux in various systems from spin-$1/2$ chains to higer-spin systems. 
This approach open new perspectives in understanding the physics 
of magnetization plateaux in higher spin and higher dimensional systems 
such as rare earth tetraborides \cite{RET}.

Our goal in this paper is to test and extend the approach developed 
by Tanaka {\it et al.}\cite{Tanaka}. First, we test the technique on 
several concrete {\it small-spin} $N$-leg ladders and $p$-merized chains, 
previously studied with the help of bosonization 
\cite{plateaux_ladd, Plateaux_pmer}. Then we extend the technique 
by combining it with ideas due to R. Shankar \cite{Shankar}, 
to account for hole doping in spin chains at non-zero magnetization. 
Unfortunately, accounting for hole doping in zero 
field has proven to be problematic for technical reasons which we are 
not going to discuss here. However, at a non-zero average magnetization 
per site the technical difficulties are lifted, which allows us to 
generalize the TTH approach to the case of doping.
After this, we use doped Hubbard  and $t-J$ chains and ladders as a 
testing ground, and re-derive some of the results previously found 
for these systems 
\cite{Cabra_irrational,Poilblanc2004,Roux_irrational,Roux_2007}. 
Then, we turn to doped higher-spin systems, which 
tend to pose a problem for the bosonization approach.

In Section \ref{sec:chain}, we illustrate the key points of 
Ref.~\onlinecite{Tanaka} by studying magnetization plateaux in an anisotropic 
spin chain with the help of coherent-state path integral technique.
In Section \ref{sec:ladder}, we extend this to a dimerized chain, 
and to a two-leg spin ladder -- and then, in section \ref{sec:extension}, 
we generalize the above to $n$-leg ladders and $p$-merized 
chains, and find that they satisfy the necessary condition 
of the Eq. (\ref{eq:condition_intro_1}).

In Section \ref{sec:holes}, we generalize the plateau condition in the 
Eq. (\ref{eq:condition_intro_1}) to account for the doping dependence. 
In a spin-$S$ system at a small hole density $\delta \ll 1$, we find the 
plateau condition to read
\ba
N \left(1-\frac{\delta}{2S}\right)(S\pm m)\in \mathbb{Z}. 
\label{Main}
\ea

The Section \ref{sec:S_3/2_chain} presents an application of the 
formalism to a doped spin-$3/2$ chain, and compares the results 
with those obtained numerically by DMRG.

In Section \ref{sec:extension_holes}, we discuss the plateau 
condition for a doped $n$-leg ladders and $p$-merized chains,  
with and without doping.

In Section \ref{sec:trimerized}, we present the results 
for a trimerized chain, and compare with DMRG results.

Finally, in Section \ref{sec:conclusions} we present the 
conclusions, possible implications and extensions of the present 
approach to higher dimensions.

\section{Anisotropic spin chain in a magnetic field}
\label{sec:chain}

\begin{figure}[t]
\includegraphics[width=0.5\textwidth]{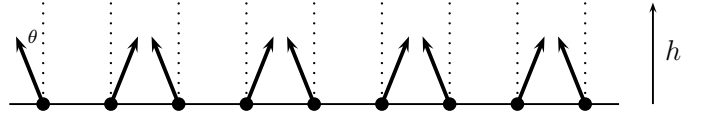}
\caption{Classical configuration for a spin chain in the presence of a magnetic field.}
\label{fig:chain}
\end{figure}

In this Section, we study the spin-$S$ nearest-neighbor 
antiferromagnetic (AF) chain with easy-plane single-ion 
anisotropy, subject to a transverse magnetic field: 
\ba
\label{eq:chain:Hamiltonian}
H=J\sum_{j}\ve{\op{S}}_{j}\cdot \ve{\op{S}}_{j+1}
 + D\sum_{j}(\op{S}^{z}_{j})^2-h\sum_{j}\op{S}^{z}_{j}, 
\ea
\normalsize
where $J$ is positive, and the magnetic field $h$ 
points in the $z$ direction, as shown in the Fig. \ref{fig:chain}.

Following the Ref.~\onlinecite{Tanaka}, we analyze the system using 
the coherent-state path integral description due to Haldane \cite{Haldane}. 
The resulting effective action comprises two terms: the first one 
is the coherent-state expectation value of the Hamiltonian, and 
the second one, dubbed the Berry phase term, corresponds to the 
solid angle swept by the spins in their imaginary time evolution.

In order to obtain an effective theory, first, we identify 
the classical ground-state configuration and the low-energy 
modes above it. Partially polarized by the magnetic field, 
the spins form a canted texture, that we parametrize as 
$\ve{S}_{j}=S \ve{n}_{j}$, with $\ve{n}_{j}$ being a unit 
vector with staggered XY components ($\phi_{j}=\frac{\pi}{a} x_{j}$): 
\ba
\ve{S}_{j}=
\left( 
  S \sin{\theta_{j}}\; \cos{\phi_{j}}, S \sin{\theta_{j}}\;\sin{\phi_{j}}, S\cos{\theta_{j}}
\right).
\ea

We parametrize the fluctuations 
around the above canted state as per 
\ba
\phi_{j}\rightarrow \frac{\pi}{a} x_{j}+\phi(x_{j})\hspace{1cm}
 \theta_{j}\rightarrow  \theta_{0}+\delta\theta(x_{j}), 
\ea
where $\theta_{0}$ is the classical ground state solution 
$\cos{\theta_{0}}=\frac{h}{2S(2J+D)}$ and $x_{j}=a j$, with $a$ the lattice constant.

Expanding up to the second order in the $\delta \theta$, 
we can write $S^{\pm}_{j}=S^{x}_{j}\pm i S^{y}_{j}$ as a function 
of $\delta\theta(x_{j})$ and $\phi(x_{j})$ and, using 
these fluctuation fields, we write an effective theory.
If we calculate the Poisson Brackets $\{ S^{z},S^{\pm}\}_{\phi,\delta\theta}$, we obtain
 $i \hbar\{ S^{z},S^{\pm}\}_{\phi,\delta\theta}=-S(\sin{\theta_{0}}
-\delta\theta\cos{\theta_{0}})\left( \pm \hbar S^{\pm}  \right)$
then  is straightforward to see that defining
\ba
\label{eq:def_PI}
a\Pi(x_{j})=-S\left[\delta \theta(x_{j}) \sin{\theta_{0}}+\frac{1}{2}(\delta \theta(x_{j}))^{2}
 \cos{\theta_{0}}\right]
\ea
as the conjugate field of $\phi$ we have the correct 
commutation relations for the spin operators
\ba
\label{eq:def_spin_z}
S^{z}_{j} &\simeq &  S\cos{\theta_{0}}+a\Pi(x_{j})\\\label{eq:def_spin_+}
S^{\pm}_{j} &\simeq &    (-1)^{j} e^{\pm i\phi(x_{j}) }
  \left[S\sin{\theta_{0}}-\frac{a\, m}{S \sin{\theta_{0}}} \Pi(x_{j}) \right.\\\nonumber
&-&\left.  \frac{a^2}{2}\frac{S^2}{S^2-m^2}\frac{1}{S \sin{\theta_{0}}}\Pi^{2}(x_{j})\right],
\ea
where $m=S\cos{\theta_{0}}$.

Following the Ref.~\onlinecite{Tanaka}, one arrives at the low-energy continuous 
effective action, corresponding to the Hamiltonian (\ref{eq:chain:Hamiltonian}): 
\ba
\nonumber
S\!\!&=&\!\!\!\int\! dx d\tau
\left\{
a\left(2J + D\right)\Pi^2
\!-\!i(\partial_{\tau}\phi)\Pi \right.\\ \label{eq:H_D_1}
&+&\left.\frac{J}{2}a(S^2-m^2)(\partial_{x}\phi)^2
 + i\left(\frac{S-m}{a}\right)(\partial_{\tau}\phi)
 \right\}.
\ea
The last term in right hand side of the Eq. (\ref{eq:H_D_1}) 
arises from the Berry phase of the individual spins in the 
Eq.~(\ref{eq:chain:Hamiltonian}). After gaussian integration 
over the field $\Pi$, the action (\ref{eq:H_D_1}) takes the form 
\small
\ba
\nonumber
S\!=\!\!\int dx d\tau
\left\{
\frac{K_{\tau}}{2}
(\partial_{\tau}\phi)^2 \!+\! \frac{K_{x}}{2}(\partial_{x}\phi)^2 
\!+\! i\left(\!\frac{S-m}{a}\!\right)\!(\partial_{\tau}\phi)
 \right\}\\\label{eq:action_chain_1}
\ea

\normalsize
with
\ba
K_{\tau}=\frac{1}{2a\left(2J+D\right)} \hspace{1cm} K_{x}=J a(S^2-m^2).
\ea

The last term in (\ref{eq:action_chain_1}) counts the winding number of the 
space-time history of the field $\phi$, defined on a covering space of a circle.
In order to understand the consequences of the topological term, it is convenient to apply a standard duality transformation to the action.
First, the phase field $\phi$ is written as $\phi=\phi_{v}+\phi_{t}$, 
where $\phi_{v}$ is a fixed field configuration containing all the vortices
$(\partial_\mu \partial_\nu-\partial_\nu\partial_\mu)\phi_{v}\neq 0$, 
and $\phi_{t}$ contains the fluctuating vortex-free part.
Next, we introduce the Hubbard-Stratonovich auxiliary vector field $J_{\mu}=(J_{\tau},J_{x})$,
and integrating by parts  we obtain
\ba
\nonumber
S&=&\int dx d\tau
\left\{\phantom{\frac12}
i\left(J_{\tau}+\left(\frac{S-m}{a}\right)\right) (\partial_{\tau}\phi_{v})  
\right. \\
&+& iJ_{x}(\partial_{x}\phi_{v})+\frac{1}{2} \frac{J^{2}_{\tau}}{K_{\tau}} +\left. \frac{1}{2} \frac{J^{2}_{x}}{K_{x}} \right.\\\nonumber
&-& \left. i \left[ \partial_{\tau} \left( J_{\tau}+ \left( \frac{S-m}{a} \right) \right)+\partial_{x}J_{x} \right] \phi_{t}
\phantom{\frac12} \right\}.
\ea
Defining $\tilde{J}_{\mu}=J_{\mu}+\delta_{\mu,0}\left(\frac{S-m}{a}\right)$ 
the vorticity-free part can be eliminated with the constraint 
$\partial_{\mu}\tilde{J}_{\mu}=0$.
Then,  the action reads
\small
\ba
\nonumber
S=\int \! dx d\tau
\left\{
i\tilde{J}_{\mu}\partial_{\mu}\phi_{v}+\frac{1}{2K_{\tau}}\left(
\tilde{J}_{\mu}-\left(\frac{S-m}{a} \right)\right)^2
\!\!+\frac{1}{2K_{x}}\tilde{J}^{2}_{x}
\right\}.
\ea
\normalsize
The constraint $\partial_{\mu}\tilde{J}_{\mu}=0$ is solved in one dimension (1D) in terms of an auxiliary field by
\ba
\tilde{J}_{\mu}=\epsilon_{\mu \nu}\partial_{\nu}\chi ,
\ea
where $\chi$ is vorticity-free. 
Integrating the first term in the action by parts, and using the substitution 
$\chi \rightarrow \tilde{\chi}+\left(\frac{S-m}{a} \right)x$,  we obtain 
%
\ba
\nonumber
S&=&\int  dx d\tau
\left\{
\frac{1}{2K_{\tau}} ( \partial_{x}\tilde{\chi})^2
+\frac{1}{2K_{x}}(\partial_{\tau}\tilde{\chi})^2 \right.\\
&+&\left. iB\left(\tilde{\chi} +\left(\frac{S-m}{a} \right)x\right)
\right\},
\ea
where 
$B=\epsilon_{\mu \nu} \partial_{\mu}\partial_{\nu} \phi_{v} =
 \sum_{j}2\pi q_{j}\delta(\tau-\tilde{\tau}_{j})\delta(x-\tilde{x}_{j})$. Here, $\tilde{\tau}_{j}$
 and $\tilde{x}_{j}$  are the time and space coordinates of the $j$-th vortex event and $q_{j}$ is the vorticity. 
 Upon summation over the vortex configurations in the partition function, 
 and then rescaling the time variable, the action is brought to the form 
\small
\ba
\nonumber
 S&=&\int dx d\tau \left\{\phantom{\frac12}
\frac{1}{2\sqrt{K_{\tau}K_{x}}} \left((
\partial_{x}\tilde{\chi})^2
+(\partial_{\tau}\tilde{\chi})^2\right)\right.\\
&+&\left.\tilde{\lambda}_{1}\cos{\left( 2\pi\left(\tilde{\chi}
+\left(\frac{S-m}{a} \right)x\right)\right)}
\phantom{\frac12} \right\}. 
\label{SG}
\ea
\normalsize
This effective action is identical to the one obtained 
in the Ref.~\onlinecite{OYA} via bosonization. 
The cosine term is commensurate for $S-m\in \mathbb{Z}$. 
The magnetic excitations are gapless for all the values of $m$ 
that satisfy $S-m\notin  \mathbb{Z} $. Then, when the condition
\ba
\label{eq:condition_1}
S-m\in  \mathbb{Z} 
\ea
 is satisfied, a plateau can occur for some range of parameters 
 if the scaling dimension of the perturbation is small enough.
More precisely, the presence of the cosine operator in the Eq. (\ref{SG}) 
is not enough to produce a gap in the spectrum. The stiffness of the 
$\tilde{\chi}$ field must be such that this operator has scaling $d$ 
dimension smaller than $2$.  Since $d$ is a decreasing function of 
$D$, we expect a plateau for large enough values of this parameter.

The condition (\ref{eq:condition_1}) is the well-known result of Oshikawa, 
Yamanaka and Affleck \cite{OYA}, which generalizes the argument due to 
Lieb, Schulz and Mattis\cite{LSM}.

A more general situation arises when $S-m=p/r$, with $p,r\in \mathbb{Z}$. 
In this case, the vortices whose winding numbers are integer multiples 
of $r$ are free of destructive interference. In this case, only vortices 
with vorticity $\pm r$ are able to condense. In the effective action, this 
is signalled by the $r$th harmonic of the cosine term in the harmonic 
expansion being commensurate.  If  relevant, this cosine operator 
gives rise to a gapped $r$-fold degenerate ground state and 
fractionalized excitations \cite{Tanaka}. 
The best-known example of such a scenario is the AF chain with a strong 
enough second-neighbor AF coupling\cite{okunishi}.  
The system is then in a gapped phase with  two degenerate spontaneously 
dimerized ground states; fractional excitations are spinons -- domain walls 
between these two ground states. In what follows, we concentrate 
on the case $S-m\in \mathbb{Z}$.

At this point, we would like to make an observation, which will prove 
useful in the Section \ref{sec:holes}, when comparing the plateau 
condition for doped systems with that at zero doping: 
According to the Eq. (\ref{eq:action_chain_1}), for each 
contribution to the partition function, the Berry phase term 
provides a weight factor 
\ba
\nonumber
e^{i(S-m)\sum_{j}\int d\tau \partial_{\tau}\phi_j}, 
\ea
where the integral $\int d\tau \partial_{\tau}\phi_j = 2\pi n_j$ 
is an integer times $2 \pi$, and the subscript $j$ labels the 
spins in the chain. Inverting the sign in front of the $S$ changes the 
imaginary exponent by $4 \pi n_j S$, which is an integer times 
 $2 \pi$ both for integer and half-integer $S$. Therefore, 
the necessary condition (\ref{eq:condition_1}) for the 
magnetization plateau may be equivalently presented as 
\ba
(S\pm m)\in \mathbb{Z}.
\label{eq:condition_equivalent}
\ea

In the following, we would like to extend the approach above to other 
one-dimensional spin systems. In the next section, we discuss the 
cases of a two-leg ladder and a dimerized spin chain. Then, in the 
Section \ref{sec:holes}, we study the magnetization plateaux in the 
presence of doping.

\section{Two-leg ladders and dimerized chains}
\label{sec:ladder}

\subsection{Two-leg spin ladder}
The formalism above can be used to study more involved spin models, 
such as spin ladders, that interpolate between one and two dimensions.
 For spin $1/2$ experimental evidence of zero magnetization plateaux has been reported for instance in
 (C$_5$H$_{12}$N)$_2$CuBr$_4$ \cite{Watson_2001}.  
In this section we study two-leg ladders with single-ion anisotropy in a 
magnetic field, and in the next section we discuss extensions to $N$-leg 
ladders and $p$-merized chains. 

Consider the following Hamiltonian
\small
\ba
\nonumber
H &=& \sum_{j}
\left\{
 J_{\shortparallel} (\ve{\op{S}}_{1,j} \cdot \ve{\op{S}}_{1,j+1}+\ve{\op{S}}_{2,j} \cdot \ve{\op{S}}_{2,j+1})\right.\\
 &+&\left. 
J_{\bot} \ve{\op{S}}_{1,j} \cdot \ve{\op{S}}_{2,j}
+D((\op{S}^{z}_{1,j})^2+(\op{S}^{z}_{2,j})^2)\right.\\\nonumber
&-&\left. h ((\op{S}^{z}_{1,j})+(\op{S}^{z}_{2,j}))
\right\},
\ea
\normalsize
where $J_{\shortparallel}$ is the antiferromagnetic coupling 
along the chain, and $J_{\bot}$ is the  inter-chain coupling.
The onsite anisotropy term is added only for completeness 
and we can recover the isotropic case by taking the $D=0$ limit.

\begin{figure}[t!]
\includegraphics[width=0.45\textwidth]{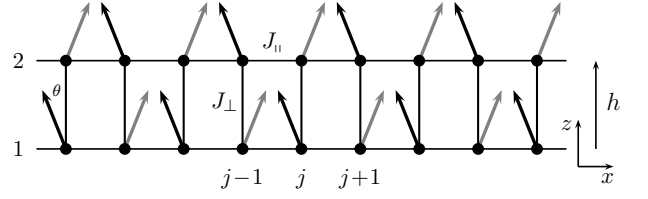}
\caption{Classical configuration for a spin ladder in the presence of a magnetic field.}
\label{fig:ladder}
\end{figure}

In the $S\rightarrow \infty$ limit we can write the energy of 
the system as a function of the angle $\theta$ (see Fig. \ref{fig:ladder}), 
with the minimum at
\ba
\cos{\theta_{0}}=\frac{h}{2S(2J_{\shortparallel}+J_{\bot}+D)}.
\ea
As before, we parametrize the fluctuations in terms 
of the fields $\delta \theta_{\alpha}$ and $\phi_{\alpha}$, 
where $\alpha=1,2$ is the chain index. 
Equations (\ref{eq:def_PI}), (\ref{eq:def_spin_z}) and (\ref{eq:def_spin_+}) 
have the same form in each chain. Using this parametrization 
in the Hamiltonian, retaining terms up to second order in the fields 
and switching to the path integral language, we obtain the action 
\small
\ba
\nonumber
S\!&=&\!\int dx d\tau \left\{
  \frac{J_{\shortparallel}}{2} a(S^2-m^2)((\partial_{x}\phi_{1})^2+(\partial_{x}\phi_{2})^2)\right.\\\nonumber
\!&+&\!\left. \frac{J_{\bot}}{2}\!\left(\frac{S^2\!-\!m^2}{a}\right)\!(\phi_{2}\!-\!\phi_{1})^2
\!+\!i \left(\frac{S\!-\!m}{a}\right)\partial_{\tau}(\phi_{1}+\phi_{2})
\right\}\\
\!&+&\!\int dx d\tau \left\{
  a  \left(2J_{\shortparallel} +J_{\bot}\frac{S^2}{S^2-m^2}+D \right)\left(\Pi_{1}^2+\Pi_{2}^2\right)\right.\\\nonumber
\!&+&\!\!\left.J_{\bot}a\left( 1-\frac{m^2}{S^2-m^2} \right)\Pi_{1}\Pi_{2}
-i \left(\Pi_{1}\partial_{\tau}\phi_{1}+\Pi_{2}\partial_{\tau}\phi_{2}\right)\right\}.
\ea
\normalsize

Integrating over the $\Pi$-fields and using 
the transformation $\vec{\varphi}=U\vec{\phi}$ with 
\ba
\label{eq:change}
\vec{\varphi}=\left(  
\begin{array}{c}
\varphi_{a}\\
\varphi_{s} 
\end{array}
 \right)
\hspace{0.5cm}
\vec{\phi}=\left(  
\begin{array}{c}
\phi_{1}\\
\phi_{2} 
\end{array}
 \right)
\hspace{0.5cm}
U=\left(  
\begin{array}{cc}
1& -1\\
\frac{1}{2} & \frac{1}{2} 
\end{array}
 \right)
\ea
we can write $S=S_{s}+S_{a}$, with

%
\ba
\nonumber
S_{s}&=&\int d\tau dx \left\{  
\frac{1}{2}K_{x}^{(s)} (\partial_{x}\varphi_{s})^{2} 
+\frac{1}{2}K_{\tau}^{(s)} (\partial_{\tau}\varphi_{s})^{2}\right.\\\nonumber
&+&\left.  i\; 2\left(\frac{S-m}{a}\right)(\partial_{\tau}\varphi_{s})
 \right\}
\ea
\normalsize
and
%
\ba
\nonumber
S_{a}&=&\int dx d\tau  \left\{  
\frac{1}{2}K_{x}^{(a)}(\partial_{x}\varphi_{a})^{2} 
+\frac{1}{2}K_{\tau}^{(a)} (\partial_{\tau}\varphi_{a})^{2}\right.\\\nonumber
&+&\left.  2J_{\bot}\left(\frac{S^{2}-m^{2}}{a}\right)\varphi_{a}^{2}
 \right\},
\ea
where 
\small
\ba
\nonumber
K_{x}^{(a)}&=&K_{x}^{(s)}=2J_{\shortparallel}a(S^{2}-m^{2})\\\nonumber
\left(K_{\tau}^{(s)}\right)^{-1}\!\! &=& \frac{a}{2}\left[4 J_{\shortparallel}+
J_{\bot}\left(1+3\frac{m^2}{S^2-m^2}\right)+2D\right]\\\nonumber
\left(K_{\tau}^{(a)}\right)^{-1}\!\! &=& \frac{a}{2}\left[4 J_{\shortparallel}+
J_{\bot}\left(3+\frac{m^2}{S^2-m^2}\right)+2D\right]
\ea
\normalsize

The action corresponding to the antisymmetric field $\varphi_{a}$
contains a mass term. In order to obtain an effective theory the field $\varphi_a$ can be 
evaluated in the saddle point solution $\varphi_{a}=0$.

In the symmetric action $S_{s}$, we use the duality transformation of 
the section \ref{sec:chain}, presenting the field $\varphi_{s}$ 
as a sum of the vortex component $\varphi_{s,v}$ and 
the vortex-free component $\varphi_{s,t}$. Upon rescaling 
the time as per $\tau\rightarrow \sqrt{K^{(s)}_{\tau}/K^{(s)}_{x}}\; \tau$, 
and following the standard steps we obtain

\small
\ba
\nonumber
 S&=&\int dx d\tau \left\{\phantom{\frac12}
\frac{1}{2\sqrt{K^{(s)}_{\tau}K^{(s)}_{x}}} \left((
\partial_{x}\tilde{\chi})^2
+(\partial_{\tau}\tilde{\chi})^2\right)\right.\\\label{eq:final_action_ladder}
&+&\left.\tilde{\lambda}_{1}\cos{\left( 2\pi\left(\tilde{\chi}
+2\left(\frac{S-m}{a} \right)x\right)\right)}
\phantom{\frac12}\!\! \right\}.
 \ea
\normalsize

Now, the cosine term is commensurate if 
\ba
\label{eq:condition_2l_lader}
2(S \pm m)\in \mathbb{Z}
\ea

Note that, even though our approach does not apply directly 
to the $m=0$ case, this conditions is nevertheless consistent 
with  the well known $m=0$ plateau for $S=1/2$.

Another important comment is due here. The factor $2$ in front 
of the Berry term is not an artifact of the transformation (\ref{eq:change}) 
because of the periodicity of the fields. Each one of the fields 
$\phi_{1}$ and $\phi_{2}$ satisfy 

\ba
\nonumber
\phi_{1}(x+L)&=&\phi_{1}(x)+2\pi n_{1}\\\nonumber
\phi_{2}(x+L)&=&\phi_{2}(x)+2\pi n_{2}
\ea
where $n_{1}$ and  $n_{1}$ are integers. Then the antisymmetric combination satisfy 
$\phi_{1}-\phi_{2}=2\pi (n_{1}-n_{2})$. 
The saddle point solution $\varphi_{a}\equiv 0$ implies
$n_{1}=n_{2}$ and then the sum $\phi_{1}+\phi_{2}$ has periodicity 
$4\pi n_{1}$. The factor $\frac12$ in the definition of $\varphi_{s}$
 is necessary for the correct periodicity. 

\vspace{0.2cm}

\subsection{Dimerized spin chain}

The dimerized spin chain can be studied in a similar way.  
We begin with the Hamiltonian

\small
\ba
\nonumber
H&=&\sum_{j}\left\{ (J+\delta J)\ve{\op{S}}_{1,j} \cdot \ve{\op{S}}_{2,j}
+(J-\delta J)\ve{\op{S}}_{2,j} \cdot \ve{\op{S}}_{1,j+1}\right.\\
&+&\left.D\left[(\op{S}^{z}_{1,j})^{2}+(\op{S}^{z}_{2,j})^{2}\right]
-h\left[\op{S}^{z}_{1,j}+\op{S}^{z}_{2,j}\right]
\right\}, 
\ea
\normalsize
and use the Eqs. (\ref{eq:def_PI}), (\ref{eq:def_spin_z}) and 
(\ref{eq:def_spin_+}) for each sublattice. We then switch to the path integral 
formalism, take the continuum limit and drop the constant terms to obtain the following action:

\ba
\nonumber
S&=&\int dx d\tau
 \left\{
J\left(\frac{S^2-m^2}{a}\right)(\phi_1-\phi_2)^2 \right.\\\nonumber
&+& \frac{a}{4}(J-\delta J)(S^2-m^2)\left[(\partial_{x}\phi_{1})^{2}+(\partial_{x}\phi_{2})^{2}\right]\\\nonumber
&+& \frac{1}{2}(J-\delta J)(S^2-m^2)(\phi_1-\phi_2)(\partial_{x}\phi_{1}+\partial_{x}\phi_{2})\\
&+&a\left( J \frac{S^2}{S^2-m^2}+D\right)(\Pi_{1}^{2}+\Pi_{2}^{2})\\\nonumber
&+&  2aJ\left(1-\frac{m^2}{S^2-m^2} \right)\Pi_{1}\Pi_{2}\\ \nonumber
&+& i\left( \frac{S-m}{a}\right)(\partial_{\tau}\phi_{1}+\partial_{\tau}\phi_{2})\\ \nonumber
&-& \left. i(\partial_{\tau}\phi_{1}\Pi_{1}+\partial_{\tau}\phi_{2}\Pi_{2})
  \right\}.
\ea
Integrating the $\Pi$-fields and making the substitution 
$\vec{\varphi}=U\vec{\phi}$ with $U$ given by (\ref{eq:change}) 
we have
\ba
\nonumber
S&=&\int dx d\tau
 \left\{
4J\left(\frac{S^2-m^2}{a}\right)\varphi_{a}^2 \right.\\\nonumber
&+& \frac{a}{2}(J-\delta J)(S^2-m^2)\left[(\partial_{x}\varphi_{s})^{2}+(\partial_{x}\varphi_{a})^{2}\right]\\\nonumber
&+& 2(J-\delta J)(S^2-m^2)\varphi_a \partial_{x}\varphi_{s}\\
&+& i2\left( \frac{S-m}{a}\right)(\partial_{\tau}\varphi_{s})\\ \nonumber
&+& \left. \frac{1}{2}\tilde{K}_{\tau}^{(s)} (\partial_{\tau}\varphi_{s})^{2}
+\frac{1}{2}\tilde{K}_{\tau}^{(a)} (\partial_{\tau}\varphi_{a})^{2}
  \right\}.
\ea

The field $\varphi_{a}$ is massive and we can use 
the saddle point solution for it $\varphi_{a}=0$. 
The action for the symmetric field results
\ba
\nonumber
S_{s}&=&\int d\tau dx \left\{  
\frac{1}{2}\tilde{K}_{x}(\partial_{x}\varphi_{s})^{2} 
+\frac{1}{2}\tilde{K}_{\tau} (\partial_{\tau}\varphi_{s})^{2}\right. \\
&+& \left.  2 i \left(\frac{S-m}{a}\right)(\partial_{\tau}\varphi_{s})
 \right\}
\ea
with $\tilde{K}_{x}=a(J-\delta J)(S^2-m^2)$ and $\tilde{K}_{\tau}=1/[a(2J+D)]$.

The action has the same form as for the two-leg ladder and, 
repeating the steps described in the preceding subsection, we obtain the necessary 
condition for the formation of magnetization plateaux
\ba
2(S\pm m)\in \mathbb{Z}.
\ea

In both cases studied in this Section, the unit cell contains 
two spins, and the effective action was written in terms of 
two fields: $\varphi_{s}$ and $\varphi_{a}$. Notice that, of these 
two, only the massless one ($\varphi_{s}$) defines the 
necessary condition for the formation of plateaux.

\section{Extensions to $N$-leg ladders and $P$-merized chains.  }
\label{sec:extension}

The arguments, presented in the preceding sections, can be easily 
extended to more complex models. As we have shown in the previous 
section, the magnetization processes of a two-leg ladder and of a 
dimerized chain are described by a single effective action of the 
massless field, that has the same form for these two different models. 
Under certain restrictions, this similarity holds 
for $N$-leg ladders and $N$-merized chains, as well. 
In the last section we have used two fields to describe the low-energy 
theory and finally only one field remains massless in our effective action.  
The extension to the $N$-leg ladder is natural and follows the same 
steps as in the two legs case but now working with $N$ different fields. 
The first $N-1$ fields are massive as in the case of $\varphi_{a}$ for 
the two legs ladder. The  effective action can be written also in terms 
of the last field, which is  given by $\varphi_{s}=\frac{1}{N}\sum_{i} \phi_{i}$, 
and the action is a straightforward generalization of the of equation 
(\ref{eq:final_action_ladder})
\small
\ba
\nonumber
 S&=&\int dx d\tau \left\{\phantom{\frac12}
\frac{2\pi^{2}}{\beta^2} \left((
\partial_{x}\tilde{\chi})^2
+(\partial_{\tau}\tilde{\chi})^2\right)\right.\\\label{eq:final_action_N}
&+&\left.\tilde{\lambda}_{1}\cos{\left( 2\pi\left(\tilde{\chi}
+N\left(\frac{S-m}{a} \right)x\right)\right)}
\phantom{\frac12} \right\},
 \ea
\normalsize
where $\beta$ depends of the microscopic parameters and  the commensurability condition is  given by
\ba
N (S \pm m)\in \mathbb{Z}.
\label{N-leg}
\ea
Again, this result contains also what is known for the 
zero magnetic field. The $N$-leg ladder at zero field 
was studied by G. Sierra \cite{Sierra} using the original 
Haldane's path integral approach, and the presence of 
a spin gap was recovered from the condition 
(\ref{N-leg}) with $m=0$.

\section{Adding holes to a single  chain}
\label{sec:holes}

There are several examples of hole-doped magnets. For instance,
 doping the $S=1$ metal oxide  Y$_2$BaNiO$_5$\cite{metal_oxides} with Ca introduces 
 hole carriers in the chains. Other examples are given by manganese 
 oxides such as La$_{1-x}$Ca$_x$MnO$_3$\cite{Dagotto_prb_manganites}.

Theoretical studies of these compounds generally depart from the double 
exchange model\cite{Dagotto_prl}. Using strong Hund's rule coupling 
between the itinerant and localized spins, an effective Hamiltonian on 
a restricted Hilbert space is introduced, where creation of a hole on a 
given site replaces the spin $S$ on this site by $S-1/2$. Such a calculation 
of the effective Hamiltonian generalizes the derivation of the 
$t-J$ model from the Hubbard model for $S=1/2$\cite{Zhang_prb}.
In this  section, we study the effect of doping on the low energy physics of 
a Heisenberg chain and the corresponding effect on the magnetization curve.


We begin with a spin chain with one spin per site in the presence of 
a magnetic field. Using the  path integral formulation, we have shown 
that the effective action is given by the Eq. (\ref{eq:H_D_1}) of the 
Section \ref{sec:chain}. Creation of a hole at a given site $j$ corresponds 
to extracting the spin from this site. To account for this, we will simply 
remove the contribution that the $j$th spin would have made to the action.
Let us introduce at each site a hole creation operator $\psi^{\dag}_{j}$, 
satisfying fermion commutation relations $\{\psi^\dag_i,\psi_j\}= \delta_{i j}$. 
Without any holes, we have just the low-energy action (\ref{eq:H_D_1}) 
of the Section \ref{sec:chain} for the pure spin system.

Important contributions to the action upon creation of a 
hole are obtained by the change in the interaction  
 between the spins, and simply replacing the Berry term 
for the spin-$S$ by that for $S - 1/2$ for the spin with the hole
  and then taking into account the contribution to the action due 
 to the hole hopping. We start by considering this very hopping term.

Let $\Omega_{1}$ and $\Omega_{2}$  be the classical spins 
at sites 1 and 2. According to the arguments of Shankar \cite{Shankar}, 
the correct hole hopping term from site 1 to site 2 would not be simply 
$-t\psi^{\dag}_{2}\psi_{1}$, as such a term would move the hole, 
but not the spin $\Omega$. At the  microscopic level, the electron 
is transferred by the operator
$\op{d}=\sum_{\sigma}\op{c}^{\dag}_{1\sigma}\op{c}_{2\sigma}$,
that moves the charge and the spin coherently from 2 to 1. 
The action of this operator on a state with the hole on site 1 is
\ba
\op{d}| o,\Omega_2 \rangle=|\Omega_2 , o \rangle
\ea
where $o$ represents a hole. In this language, 
the correct matrix element for the process is
$-t\langle \Omega_{1} | \Omega_{2} \rangle $. 
Thus, the correct hopping term is 
$-t\langle \Omega_{1} | \Omega_{2} \rangle\; \psi^{\dag}_{2}\psi_{1}$, where
\ba
\langle \Omega_{1}|\Omega_{2} \rangle =\left(\cos{\frac{\theta_{1}}{2}}\cos{\frac{\theta_{2}}{2}}+
e^{i(\phi_{2}-\phi_{1})}\sin{\frac{\theta_{1}}{2}}\sin{\frac{\theta_{2}}{2}}\right)^{2S}
\ea
Notice that, in the $h=0$ case, the classical configuration of the spins is antiparalel and 
the overlap of coherent states vanishes.
Then, in the zero magnetic field case, the hoping amplitude between nearest neighbors is a fluctuating variable with 
zero average, a very difficult problem to study. This is the main reason for the original problem 
on hole doping\cite{Shankar} to concentrate in a model where doping between second neighbors was the dominant effect.
In our case, the non zero magnetization implies a non zero overlap between neighboring coherent states allowing 
for a consistent study of first neighbors hoping problem.

Finally, the hopping of holes is described by the following Hamiltonian
\ba
H_{hopp}=-t\sum_{j} \langle \Omega_{j} | \Omega_{j+1} \rangle\; \psi^{\dag}_{j+1}\psi_{j}+\mbox{h.c}.
\ea
If we expand $\theta$ around the classical energy 
minimum value $\theta(x)=\theta_{0}+\delta \theta(x)$, 
we must retain the terms up to order $a$, since the continuum limit 
$\psi_{j}\rightarrow \sqrt{a}\psi(x)$ for the fermion operators involves 
an extra factor of $a$. Using the Eq. (\ref{eq:def_PI}) and retaining 
terms up to the first order in $a$, we find  
\ba
\langle \Omega_{j}|\Omega_{j+1} \rangle \simeq \left(\frac{m}{S}\right)^{2S}e^{\left( 2a \frac{S}{m}\Pi(x)-ia\frac{S}{m}(S-m)\partial_{x}\phi(x)\right)}
\ea
then
%
\small
\ba
\nonumber
H_{hopp}\!\!\!&\!\simeq \!\!\! &\!\! -t\sum_{j}\! \left(\frac{m}{S}\right)^{2S}\!\!\!\! \psi^{\dag}_{j+1}\psi_{j}\!\left(\!1\!+\!  2a \frac{S}{m}\Pi(x)\!-\!ia\frac{S}{m}(S\!-\!m)\partial_{x}\phi(x)\!\right)\\
& +& \mbox{h.c}.
\ea
\normalsize

Since the sought long-distance physics involves only the states near 
the Fermi surface, we linearize the theory around $\pm k_{F}$ 
\ba
\psi_{j}=e^{ik_{F} a j}\psi_{R,j}+e^{-ik_{F} a j}\psi_{L,j}, 
\ea
to obtain $H_{hopp}=H_{free}+\delta H$, where
\ba
\nonumber
H_{free}&=&\sum_{j}
-t\cos{k_{F}a}
\left[
\psi^{\dag}_{R,j+1}\psi_{R,j}  +  \psi^{\dag}_{R,1}\psi_{R,j+1}  \right.\\ \nonumber
&+&\left.\psi^{\dag}_{L,j+1}\psi_{L,j}  +  \psi^{\dag}_{L,1}\psi_{L,j+1}
\right]\\
&+&it\sin{k_{F}a}
\left[
\psi^{\dag}_{R,j+1}\psi_{R,j}  -  \psi^{\dag}_{R,1}\psi_{R,j+1}  \right.\\ \nonumber
&-&\left. \psi^{\dag}_{L,j+1}\psi_{L,j}  +  \psi^{\dag}_{L,1}\psi_{L,j+1}
\right]
\ea
and 
\small
\ba
\nonumber
\delta H \!\!& \!\!=\!&\!\!\! -t\left(\frac{m}{S}\right)^{2S}
\sum_{j}\left\{
4a\frac{S}{m}\cos{(k_{F}a)}\Pi(x)\left( \psi^{\dag}_{R,j}\psi_{R,j}  + \psi^{\dag}_{L,j}\psi_{L,j} \right)
\right.\\
&-&\!\! \left.  2a\frac{S}{m}(S-m)\sin{(k_{F}a)}\partial_{x}\phi(x)\left( \psi^{\dag}_{R,j}\psi_{R,j}  - \psi^{\dag}_{L,j}\psi_{L,j} \right)\right\}
\ea
\normalsize
Taking the continuum limit $\psi_{j}\rightarrow \sqrt{a}\psi(x)$ 
and setting the Fermi energy to zero, we obtain
\small
\begin{widetext}
\ba
\nonumber
H_{hopp}\!\!&=&\!\!\!\int \! dx
\left\{  2at
\left(
             \frac{m}{S}\right)^{2S}\!\!\! \sin{(k_{F}a)}  \left[  \psi^{\dag}_{R}(x)\left(\! -i\partial_{x} \!+\!\frac{S(S-m)}{m}\partial_{x}\phi\right)\psi_{R}(x)
           -   \psi^{\dag}_{L}(x)\left( -i\partial_{x} +\frac{S(S-m)}{m}\partial_{x}\phi\right)\psi_{L}(x)  \right] 
\right.\\
&-&\left.
4atSm^{2S-1}\cos{(k_{F}a)}\Pi(x)\left( \psi^{\dag}_{R}(x)\psi_{R}(x)+\psi^{\dag}_{L}(x)\psi_{L}(x)\right)
\right\}
\ea
\end{widetext}
\normalsize
Since the linearized theory has an infinite number of particles in the ground 
state, we shall introduce normal ordering, to correctly define the theory: 
\ba
\psi^{\dag}(x)\psi(x)=\delta+:\psi^{\dag}_{R}(x)\psi_{R}(x)+\psi^{\dag}_{L}(x)\psi_{L}(x):
\label{dis-cont}
\ea

Now, we account for hole doping. For $S=1/2$, we begin with the 
action without holes, and remove the Berry phase term at each hole site. 
For larger values of $S$, we have more than one electron on each site and a hole corresponds to 
a spin $[S-1/2]$ impurity in the 
spin-$S$ host\cite{Dagotto_prb,Dagotto_prl,Sobiella_prl}. 
In other words, we apply the projection operator
\ba
\label{eq:proy_1_s12}
\op{\mathcal{P}}_{j}=1-\frac{\psi_{j}^{\dag}\psi_{j} }{2S}
\ea
to the Berry phase term in the action if the hole 
was created on site $j$. We must do the same for the spin-spin 
interaction part using the projector.
\ba
\label{eq:proy_2_s12}
\op{\mathcal{P}}_{i,j}=(1-\frac{\psi_{i}^{\dag}\psi_{i}}{2S})(1-\frac{\psi_{j}^{\dag}\psi_{j}}{2S} ).
\ea
Note that this is somehow a different scenario than the one proposed by Shankar\cite{Shankar}
 where an entire spin $S$ would hope from one site to another.
 Our approach is more appropriate to cope with the experimental situation we
mention below. The modification of our approach to handle Shankar scenario, or any intermediate value of the spin 
hoping between $S$ and $1/2$
 is straightforward.

The following step is to linearize the theory near 
$\pm k_{F}$, as we did in the hopping term, to obtain 
\ba
\label{eq:proy_3_s12}
 \op{\mathcal{P}}_{j} & \simeq & 1-\frac{\psi_{j}^{\dag}\psi_{j}}{2S}\\\nonumber
&= & 1-\frac{\delta}{2S}-\frac{1}{2S}:\psi_{R,j}^{\dag}\psi_{R,j}+\psi_{L,j}^{\dag}\psi_{L,j}:
\ea
where $\delta$ is the noninteracting ground state expectation value.
Then, in the presence of doping, the equation (\ref{eq:H_D_1}) reads
\begin{widetext}
 
\ba
\nonumber
S_{SM}&=&\int dx d\tau
\left\{\phantom{\frac12}
\frac{J}{2}\left(1-\frac{\delta}{2S}\right)^{2}a(S^2-m^2)(\partial_{x}\phi)^2+
a\left(1-\frac{\delta}{2S}\right)^{2}\left(2J+D\right)\Pi^2
\phantom{\frac12} \right\}\\
&+& i\int dx d\tau \left\{ \left(1-\frac{\delta}{2S}\right)\left(\frac{S-m}{a}\right)(\partial_{\tau}\phi)
-\left(1-\frac{\delta}{2S}\right)(\partial_{\tau}\phi)\Pi \right.\\\nonumber
&-&\left.\frac{1}{2S}\left(\frac{S-m}{a}\right)(\partial_{\tau}\phi)(:\psi_{R,j}^{\dag}\psi_{R,j}+\psi_{L,j}^{\dag}\psi_{L,j}:)  \right\}.
\ea

\end{widetext}
Certain terms have been dropped here as irrelevant in the renormalization 
group (RG) sense, such as products of normal-ordered fermion operators 
and bilinears of spin phase operators. 
As a result, the total action is given by $S_{SM}+S_{hopp}$.
Integrating out the $\Pi$ field, we obtain
\small
\ba
\nonumber
\mathcal{S}_{eff} \!\! &=& \!\! \int dx d\tau
\left\{\phantom{\frac12}
\frac{1}{2}K_{x}(\partial_{x}\phi)^2 + 
\frac{1}{2}K_{t}(\partial_{\tau}\phi)^2\right.\\
 \!\! &+& \!\! \left. i\left(1-\frac{\delta}{2S}\right)\left(\frac{S-m}{a}\right)(\partial_{\tau}\phi)\right\}\\\nonumber
\!\! &+&\!\! ig_{1}\!\int \! dx d\tau
\left\{  
   \bar{\Psi}(x)\sigma_{3}\left( \partial_{x}\! +\! i\frac{S(S-m)}{m}\partial_{x}\phi\right) \! \Psi(x) \! \right\}  \\\nonumber
\!\! &-&\!\! \int dx d\tau
\left\{  
   \bar{\Psi}(x) \mathbb{I}  \left( \partial_{\tau} -ig_{2}(\partial_{\tau}\phi)\right)\Psi(x)  \right\},
\ea
\normalsize
where
\ba
\nonumber
\frac{1}{2}K_{x} \!\! &=& \!\! a\frac{J}{2}\left(1-\frac{\delta}{2S}\right)^{2}(S^2-m^2)\\
\frac{1}{2}K_{t} \!\! &=& \!\! \frac{1}{4a\left(2J+D\right)}\\\nonumber
g_{1} \!\! &=& \!\! 2at\left( \frac{m}{S}\right)^{2S}  \sin{(k_{F}a)}\\\nonumber
g_{2} \!\!\! &=& \!\!\! \left[\frac{1}{2S}\!\left(\!\frac{S-m}{a}\!\right)\!+\!  \frac{2atSm^{2S-1}\cos{(k_{F}a)}}{\left(1\!-\!\frac{\delta}{2S}\right)\left(2J+D\right)}\!    \right] \\\nonumber
\Psi(x)&=&\left( 
\begin{array}{c}
 \psi_{R}\\
\psi_{L}
\end{array}
\right)
\hspace{0.8cm}
\bar{\Psi}(x)=\left(  \bar{\psi}_{R},\bar{\psi}_{L}\right). 
\ea
Upon rescaling time in the fermionic part, we find  
\ba
\label{eq:SSS}
\mathcal{S}_{eff}&=&\mathcal{S}_{\phi}+\mathcal{S}_{F}
\ea
with
\small
\ba
\nonumber
\mathcal{S}_{\phi}&=&\int dx d\tau
\left\{\phantom{\frac12}
\frac{1}{2}K_{x}(\partial_{x}\phi)^2 + 
\frac{1}{2}K_{t}(\partial_{\tau}\phi)^2 \right. \\ \ \nonumber
 &+& \left. i\left(1-\frac{\delta}{2S}\right)\left(\frac{S-m}{a}\right)(\partial_{\tau}\phi)\right\}\\\nonumber
\mathcal{S}_{F}&=&i\int dx d\tau
\left\{  
   \bar{\Psi}(x)\sigma_{3}\left( \partial_{x} +i\frac{S(S-m)}{m}\partial_{x}\phi\right)\Psi(x)  \right\}  \\\nonumber
&-&\int dx d\tau
\left\{  
   \bar{\Psi}(x) \mathbb{I}  \left( \partial_{\tau} +i\frac{S(S-m)}{m}(\partial_{\tau}\phi)\right)\Psi(x)  \right\}\\\nonumber
&+&i\int dx d\tau
\left\{  
   \bar{\Psi}(x) \mathbb{I}  \left( g_{2} +\frac{S(S-m)}{m}(\partial_{\tau}\phi)\right)\Psi(x)  \right\}.
\ea
\normalsize
Now,  we write $\phi=\phi_{v}+\phi_{t}$ and then we eliminate the $\phi_{t}$ field  from the two
first terms of 
$\mathcal{S}_{F}$ via the change $\Psi\rightarrow e^{-i\frac{S(S-m)}{m}}\Psi$, and
by an appropriate rescaling, this effective action can be rewritten in a more compact form:
\ba
\mathcal{S}_{F}=\int \! dx d\tau   \bar{\Psi} \left[- \gamma_{\mu} (\partial_{\mu} + i A_{\mu} + i \tilde{A}_{\mu}) \right] \Psi
\ea
with
\ba
A_{\mu} =\frac{S(S-m)}{m}\partial_{\mu}\phi_{v} 
\ea
and
\ba
\nonumber
\tilde{A}_{0} = \left(g_{2}+\frac{S(S-m)}{m}\partial_{\mu}\right)\partial_{\tau}(\phi_{v}+\phi_{t})~;~
 \tilde{A}_{1} =0
\ea
\ba
\gamma_{0} =
\left(
\begin{array}{cc}
 0&-i\\
 i&0
\end{array}
\right)
 ~;~ 
\gamma_{1} =
\left(
\begin{array}{cc}
 0&1\\
 1&0
\end{array}
\right).
\ea

As we have retained only quadratic terms, we can integrate out the fermionic  
degrees of freedom. The result is just the determinant of the fermion's kernel. At this point 
there is a mathematical observation which is important to stress. The Atiyah-Singer 
Index theorem stress that the fermionic determinant is non zero only when 
 the gauge fields $A$ and $\tilde{A}$ have zero 
total magnetic flux\cite{Shankar}. 
It can be shown that this condition is automatically satisfied for the field 
$\tilde{A}$. For the field  $A$, the Index theorem then imposes a global 
constraint that the field $\phi$ have the total vorticity equal to zero, which is nothing else than the 
charge neutrality of the vortex gas realized by the field $\phi_{v}$.

Once we have integrated out the fermions, we are left with an 
effective action, which depends only of the scalar field $\phi$.
It is  basically given by $\mathcal{S}_{\phi}$ in Equation (\ref{eq:SSS}) corrected by unimportant gauge invariant 
counter-terms arising from the fermionic determinant. 
We then perform the same steps as above, in particular the duality transformation, and get an effective action of the
kind:
\small
\ba
\nonumber
\mathcal{S}_{\phi}&=&\int dx d\tau
\left\{\phantom{\frac12}
\frac{1}{2K_{\tau}} ( \partial_{x}\tilde{\chi})^2
+\frac{1}{2K_{x}}(\partial_{\tau}\tilde{\chi})^2 \right. \\ \nonumber
&+& \left.\lambda_{1}\cos{\left[ 2\pi\left(\tilde{\chi}
+\left(1-\frac{\delta}{2S}\right)\left(\frac{S-m}{a} \right)x\right)\right]}
\phantom{\frac12}\!\!\! \right\}.\\
\ea
\normalsize

In the action above, the cosine term is commensurate 
when the following condition is satisfied 
\ba
\label{eq:condition}
\left(1-\frac{\delta}{2S}\right)\left(\phantom{\frac12}\!\!\!\!S-m \right)\in  \mathbb{Z}.
\ea

Let us now compare the plateau condition (\ref{eq:condition}) 
for the doped chain to the zero-doping condition 
(\ref{eq:condition_equivalent}), namely 
\ba
(S\pm m)\in \mathbb{Z}. 
\nonumber
\ea
The latter emerged from rewriting the Berry phase term 
in two different yet equivalent ways. Now, if we follow the 
same arguments and introduce the projectors in the Eqs. 
(\ref{eq:proy_1_s12}) and (\ref{eq:proy_2_s12}), 
we obtain the two following conditions
\ba
\left(1-\frac{\delta}{2S}\right)(S\pm m)\in \mathbb{Z}. 
\label{cond_dop}
\ea
Notice that, for $\delta \neq 0$, the two signs in the condition above 
no longer define the same set of plateau magnetization values \cite{comment1}. 
This inequivalence appeared as a result of passing to the continuum 
limit and implementing the normal ordering (\ref{dis-cont}). The linear 
approximation can be implemented provided interference terms 
due to the Berry phase are properly taken into account, 
{\it i.e.} if there is a value of doping and magnetization, 
for which vortex configurations are not suppressed by the 
Berry phase, this should also be true in the linearized theory. 
Then, in the linearized theory, if one of the conditions in 
(\ref{cond_dop}) is satisfied, it must manifest itself in the 
effective action for the scalar field $\phi$ by a commensurate 
cosine operator.  
Upon infinitesimal doping, each plateau magnetization 
value that meets the zero-doping plateau condition 
\ba
(S\pm m)\in \mathbb{Z}, 
\ea
gives rise to two different plateau magnetization values, 
according to the Eq. (\ref{cond_dop}). However, it is the 
microscopic details of the model that eventually determine  
whether the plateau is indeed realized at both 
of these values, only one of them -- or neither. 


%
\section{Magnetization plateaux of an anisotropic $S=3/2$ spin chain. }
\label{sec:S_3/2_chain}

Now, we use the results presented in the preceding sections to study 
the magnetization plateaux of a simple spin chain. The $S=3/2$ chain 
with on-site anisotropy has been studied numerically by Okamoto 
\cite{Okamoto} and Sakai and Takahashi \cite{Sakai}. 
 At $D=0$, there are no plateaux in the magnetization curve; as $D$ 
 is increased beyond a critical values $D_{c}$, a plateau appears at $m_{sat}/3$, where $m_{sat}$ denotes the saturation value.
 With the formalism developped in the sections above, we qualitatively recover 
 these results, and show that the critical value $D_{c}$ decreases upon doping. 
 Indeed, as mentioned above, the stiffness of the field $\tilde{\chi}$ must be 
 such that the cosine operator in ({\ref{SG}) be relevant in the renormalization 
 group sense. In the case $\delta=0$, our computation shows that this is achieved 
 for values of $D$ greater than $D_{c}=0.467$, a value to be compared 
 with the result of Okamoto and Kitazawa ($D_{c1}=0.387$) .

In the presence of doping, the Hamiltonian contains lattice $S = 3/2$ spins and spin-1 mobile 
``holes'', and can be obtained from a Kondo lattice model~\cite{Dagotto_prb}. 
It has the form 
\begin{equation}\label{tj32.eq}
{\cal H} = \sum_{\langle ij\rangle} -t \hat{T} + 
J \vec{S}_i \cdot \vec{S}_j + \sum_i D (S_i^z)^2 , 
\end{equation}
where the kinetic energy term acts as per 
\begin{eqnarray}
\hat{T} |3/2,m\rangle|1,m'\rangle & = &  A |1,m+1/2\rangle|3/2,m'-1/2\rangle \nonumber \\ &+& B |1,m-1/2\rangle|3/2,m'+1/2\rangle
\end{eqnarray}
with $A=\sqrt{(2-m')(3/2-m)}/3$ and $B=\sqrt{(2+m')(3/2+m)}/3$. 
In other words, the local magnetization can only change by 1/2 
since it comes originally from the hopping of spin-1/2 electrons. 

Upon doping, we expect the plateau to split into two. The necessary condition 
for the formation of magnetic plateaux in the spin-$3/2$ chain is 
given by
\ba
(1-\frac{\delta}{3})(\frac{3}{2}-m)\in \mathbb{Z}
\ea
For a small enough $\delta$ we obtain the following possible plateaux
\ba
\tilde{m}_{\pm}=\frac{1}{3}(1\pm \delta)\;\; \hbox{and}\;\; \tilde{m}_{s}=1-\frac{\delta}{3}
\ea
where $\tilde{m}=(1-\frac{\delta}{2S})\frac{m}{S}$.

In order to check these predictions, we have performed numerical 
simulations using  the powerful DMRG algorithm~\cite{dmrg} for 
various dopings at a fixed large $D/J=5$, and with open boundary 
conditions (OBC) for system lengths up to $64$. Typically, we kept 
up to 1200 states, which is sufficient to have a discarded weight 
smaller than $10^{-12}$.

The DMRG results are showed in Fig.~\ref{fig:plateau_DMRG_3_2}. 
Without doping, the large on-site anisotropy $D$ stabilizes a wide 
magnetization plateau at $M_{s}/3$ in agreement with previous 
studies~\cite{Okamoto,Sakai}. Now, doping with $S=1$ impurities 
splits this plateau, in perfect agreement with our prediction. 

\begin{figure}[t!]
\includegraphics[width=0.45\textwidth,clip]{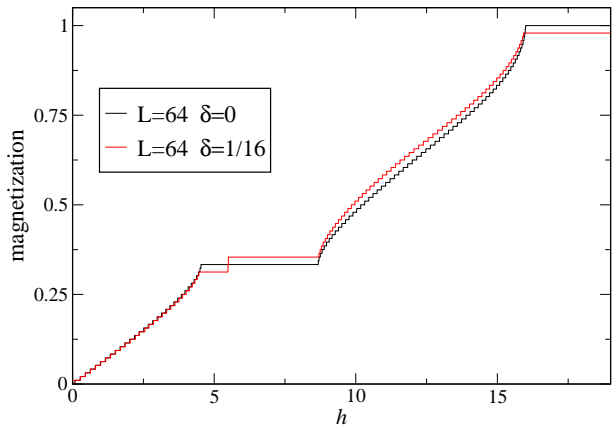}
\caption{(Color online) Magnetization curve ($\tilde{m}$) for the $S=3/2$ chain corresponding to $D/J=5$ obtained with DMRG.}
\label{fig:plateau_DMRG_3_2}
\end{figure}
%


\section{Holes in $N$-leg ladders and $p$-merized chains}
\label{sec:extension_holes}

In more complex models such as $N$-leg ladders and $p$-merized chains,
the contribution of doping can be traced similarly to how it was done in the 
Section \ref{sec:holes}. Here, we briefly discuss the approach for  the case 
of  ladders. The $p$-merized chain follows the same steps. First we note that
the hopping constant must be replaced by 
$t_{\alpha,\beta}=-t\langle \Omega_{\alpha,j}|\Omega_{\beta,j+1}\rangle$, 
where the indices $\alpha$ and $\beta$ label the chains of the ladder. 
We need to define $N$ kinds of holes corresponding to each chain. 
Then we have two kinds of hopping terms, the hopping in each chain 
($t_{\alpha,\alpha}$) and the interchain hopping
 ($t_{\alpha,\beta}$, with $\alpha\neq\beta$). 
 Straightforward calculations give
\small
\ba
\!\!t_{\alpha,\alpha}&\simeq& t
 \left(\frac{m}{S}\right)^{2S}\!\!\exp{\left[ 2a \frac{S}{m}\Pi_{\alpha}
-ia\frac{S}{m}(S-m)\partial_{x}\phi_{\alpha}\right]}\\\nonumber
\!\!t_{\alpha,\beta}&\simeq& t \left(\frac{m}{S}\right)^{2S}
\!\!\exp{ \left[ 2a \frac{S}{m}(1-i(\frac{\phi_{\alpha}-\phi_{\beta}}{2}))(\Pi_{\alpha}-\Pi_{\beta})\right]}\\
&\times&\exp{\left[i2\frac{S}{m}(S-m)(\frac{\phi_{\alpha}-\phi_{\beta}}{2})\right]}.
\ea
\normalsize

Now, we can write the kinetic Hamiltonian for the fermions and linearize the spectrum as before.
 In the spin part and
the Berry phase contribution we must insert the corresponding projectors 
$1-\frac{\psi^{\dag}_{\alpha}\psi_{\alpha}}{2S}$ and 
$(1-\frac{\psi^{\dag}_{\alpha}\psi_{\alpha}}{2S})
(1-\frac{\psi^{\dag}_{\beta}\psi_{\beta}}{2S})$. The procedure is a straightforward extension of the spin chain
case: we must integrate over the $\Pi_{\alpha}$ fields and use  the Hubbard-Stratonovich auxiliary fields as in the case of the spin chain, decoupling each field $\phi_\alpha$ as
 $\phi_\alpha=\phi_{\alpha,v}+\phi_{\alpha,t}$. The first $N-1$ fields are massive and can be
evaluated in the saddle point solution. Once we have integrated out the fermions,
 we finally obtain the following result for the effective action
\small
\ba
\nonumber
S_{eff}^{ladder}\!\!\!&=&\!\!\!\int dx d\tau
\left\{\!\!\!\!\phantom{\frac12}
\frac{1}{2K_{\tau}} ( \partial_{x}\tilde{\chi})^2
+\frac{1}{2K_{x}}(\partial_{\tau}\tilde{\chi})^2 \right. \\ \nonumber
\!\!\!&+& \!\!\!\left.\lambda_{1}\cos{\left( 2\pi\left(\tilde{\chi}
+N\left(1\!-\!\frac{\delta}{2S}\right)\left(\frac{S\!-\!m}{a} \right)x\right)\right)}
\phantom{\frac12}\!\!\! \right\}.\\
\ea
\normalsize

The final condition for the formation of magnetization plateaux reads
\ba
\label{eq:condition_Nlegs}
N\left(1-\frac{\delta}{2S}\right)\left(S\pm m \right) \in \mathbb{Z}.
\ea
The calculation for the $N$-merized chain is straightforward and gives the condition (\ref{eq:condition_Nlegs}).


For the case of the two-leg ladder, plateaux at irrational values controlled by doping have been predicted by bosonization \cite{Cabra_irrational}
 and numerically supported by numerical results \cite{Roux_irrational,Roux_2007}. In our approach the condition for the formation of magnetization plateaux in
a $S=1/2$ two legs ladder with doping reads
\ba
2(1-\delta)(S-m) \in \mathbb{Z}
\ea
where the magnetization per site takes values between $-S\leq m\leq S$.
 To compare with the results in Refs.\onlinecite{Cabra_irrational} and
 \onlinecite{Roux_irrational,Roux_2007} we must properly normalize the magnetization as $\tilde{m}=\frac{(1-\delta)}{S}m$, and then, for $S=1/2$ the condition reads
\ba
 1-\delta \pm \tilde{m} \in \mathbb{Z}
\ea
which is identical to the condition used in Ref.~\onlinecite{Roux_irrational} based on the OYA criterion~\cite{OYA}.
For small $\delta$ we expect possible plateaux at $\tilde{m}=1-\delta$ and $\tilde{m}=\delta$, which has been confirmed numerically with DMRG (See Fig. 3 of Ref.~\onlinecite{Roux_2007}).


\section{Trimerized chain}
\label{sec:trimerized}
In this section we study a $S=1/2$ trimerized chain in a magnetic field.
Although the main goal of this section is to test our procedure  
developped in the previous sections against the numerical results, 
obtained by means of DMRG simulations, the study of this model 
in the presence of doping is interesting by itself\cite{Cabra_phys_lett_a}. For instance, 
the antiferromagnetic $S=1/2$ trimerized chain has experimental 
realizations such as the synthesized copper hydroxydiphosphate 
Cu$_3$(P$_2$O$_6$OH)$_2$, where a $1/3$ magnetization 
plateau has been observed~\cite{Hase_2006}.

\begin{figure}[t!]
\includegraphics[width=0.45\textwidth,clip]{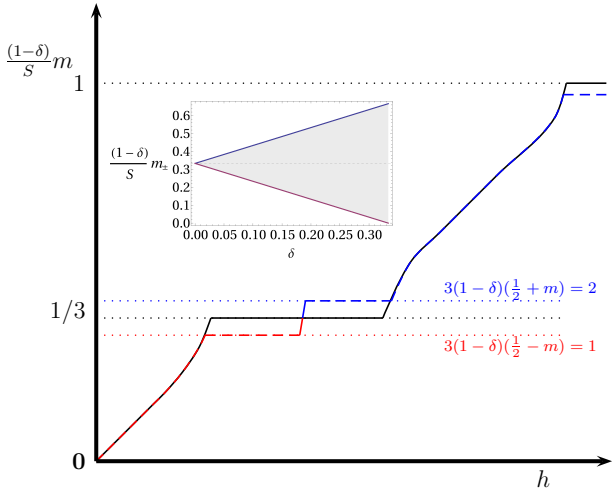}
\caption{(Color online) Schematic magnetization curve for the $S=1/2$ trimerized chain. 
The solid line corresponds to $\delta=0$  and the dashed red(blue) one 
to $\delta\neq 0$. In the inset we show the position of the possible 
magnetization plateaux as a function of $\delta$. }
\label{fig:schematic_plateau}
\end{figure}

The Hamiltonian of the $S=1/2$ trimerized chain can be written as
\ba
\nonumber
H&=&J\sum_{j} (\vec{\op{S}}_{j,1}\cdot \vec{\op{S}}_{j,2}+\vec{\op{S}}_{j,2}\cdot \vec{\op{S}}_{j,3})\\ 
&+& \frac{\gamma J}{2}\sum_{j}(\vec{\op{S}}_{j-1,3}\cdot \vec{\op{S}}_{j,1}+\vec{\op{S}}_{j,3}\cdot \vec{\op{S}}_{j+1,1})\\\nonumber
&-&h\sum_{j}\sum_{\alpha=1}^{3}\op{S}^{z}_{j,\alpha}
\ea
where $J$ is the intra-trimer coupling and $\gamma J$ the inter-trimer one. 
The first subscript $j$ labels a trimer, while the second subscript labels the 
spins within the trimer. We proceed as in the case of the dimerized chain 
 -- except that now we work with three fields corresponding to the three 
 different spins in each trimer. Eventually, only one field remains massless
 in the effective action, which is a straightforward generalization of the 
Eq.~(\ref{eq:action_chain_1}). 
  The commensurability conditions are given by the 
  Eq. (\ref{eq:condition_Nlegs}) with $N=3$ and $S=1/2$:
\ba
3(1-\delta)(\frac12\pm m) \in \mathbb{Z}
\ea
where we take $-\frac12 \leq m \leq \frac12$. For $\delta=0$ a magnetization plateau is expected at $m/m_{sat}=1/3$.
In the presence of doping, the plateau is expected 
to split into  two different plateaux, located at
\ba
\frac{m_{\pm}}{S}=\frac{1}{3}\left( \frac{1\pm 3\delta}{1-\delta}\right)
\ea

As usual, a simple way to introduce doping is to consider a $t-J$ hamiltonian: 
\begin{eqnarray}\label{tj.eq}
H &= &\sum_{\langle ij\rangle} J_{ij}(\vec{\op{S}}_{i}\cdot \vec{\op{S}}_{j}-\frac{1}{4}n_i n_j) - \sum_{\langle ij\rangle,\sigma} t_{ij} (c^\dagger_{i\sigma} c_{i\sigma} + h.c)\nonumber\\ 
&-&h\sum_{j}\sum_{\alpha=1}^{3}\op{S}^{z}_{j,\alpha}
\end{eqnarray}
where the double occupancy is forbidden on each site, the magnetic exchange 
and hopping amplitudes are equal, respectively, to $J$ and $t$ for intra-trimer 
bonds -- and to $\gamma J$ and $\sqrt{\gamma}\, t$ for inter-trimer bonds. 
Note that we have used an inter-trimer hopping amplitude of $t'=\sqrt{\gamma}t$ 
in order to be consistent with the magnetic exchange anisotropy $(t'/t)^2=\gamma$. 

In the Fig.~\ref{fig:schematic_plateau}, we show a schematic magnetization 
curve for the $S=1/2$ trimerized chain with the above parameters. We plot 
the normalized magnetization $\tilde{m}=\frac{(1-\delta)}{S}m$ in order to 
compare with the numerical results obtained by DMRG. In terms of
$\tilde{m}$, the two plateaux are predicted to satisfy 
\ba
\tilde{m}_{\pm}=\frac{1}{3}\pm \delta
\ea

\begin{figure}[t!]
\includegraphics[width=0.45\textwidth]{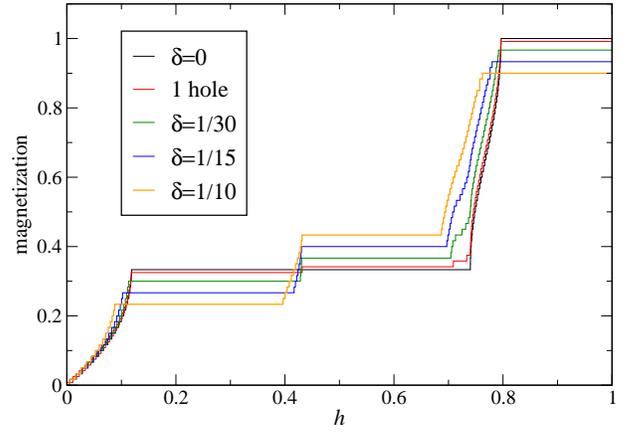}
\caption{(Color online) Magnetization $\tilde{m}$ for the $S=1/2$ trimerized chain with $\gamma=1/4$ and $J/t=0.5$,
 obtained by DMRG simulations for different values of the filling on chain of length $L=120$.
  A splitting (proportional to doping) in the magnetization plateau 
is clearly observed for non-zero doping. }
\label{fig:plateau_DMRG}
\end{figure}

The DMRG results are shown in the Fig.~\ref{fig:plateau_DMRG} for a chain
of length $L=120$ with OBC. Parameters of the $t-J$ model are $J/t=0.5$ 
and an anisotropy $\gamma=1/4$.   
Typically,
we keep up to 1200 states, which is sufficient to have a discarded
weight smaller than $10^{-9}$. In the absence of doping, our choice 
of strong anisotropy leads to a wide plateau at $m_{sat}/3$. For a finite
doping of the order of $10\%$, our numerical data are in perfect
agreement with our prediction: the plateau is split and the splitting 
is simply proportional to doping. Between the plateaux, the magnetization 
curve is smooth (the steps are only finite-size effects). However, 
for very small doping, we observe a strange shape of the magnetization 
curve just above the upper split plateau. This phenomenon can already 
be observed for a chain doped with a single hole, and it is likely to stem 
from the existence of a bound state between a hole and a polarized trimer 
(``magnon''). The same mechanism is at work for instance in lightly doped 
two-leg ladders~\cite{Troyer1996,Poilblanc2004}, where a hole pair-magnon 
bound state emerges, leading to an irrational magnetization 
plateau~\cite{Roux_irrational}. The formation of such bound states is generic 
in $t-J$ models due to the Nagaoka effect~\cite{Nagaoka}: holes gain kinetic energy 
in a ferromagnetic environment. We have not investigated in detail  
how this bound-state will modify the magnetization curve in our case, 
but our data clearly show that the magnetization curve has two different 
regimes in the upper part.

\section{Summary and conclusions}
\label{sec:conclusions}

Systems with strong easy-axis anisotropy are likely to show plateaux
in the magnetization curve even at the classical level. One
peculiarity of such plateaux is that the spin configuration must be collinear, 
{\it i.e.} all the spins of the system must be pointing in the same direction 
\cite{LhuiMis}. In the present work, we have studied the plateaux that 
are intrinsically quantum-mechanical: the magnetization curve of 
the corresponding classical system would remain a straight line all 
the way to the saturation point. This observation urges one to look 
for a theory, that would clearly identify the quantum-mechanical effects 
behind the plateaux. And this was precisely the achievement of Tanaka, Totsuka and Hu~\cite{Tanaka}. 
A key advantage of their approach versus the abelian bosonization 
is the applicability of the former in more than one spatial dimension. 

We have seen that, within the path integral approach, all the information  
relevant for the presence of a plateau is encapsulated in the Berry phase 
term. The topological nature of the Berry phase and its expression 
in terms of quantized vorticity play a crucial role here. Moreover, even 
though the path integral approach is based on the coherent-state 
description, developped for higher spins $S \gg 1$, the topological 
(quantized) nature of the Berry phase serves as a {\it protection } 
that makes it exact even for spin-$1/2$. To be more precise -- given 
a value of the plateau magnetization, the critical values of the microscopic 
parameters for the opening of the plateau are sensitive to the choice 
of a cut-off procedure, and subject to $1/S$ corrections, which renders 
their accurate calculation difficult. However, the values of the plateau 
magnetization itself, defined by the necessary condition above, 
are exact.

In this paper, we have extended the TTH approach to the presence 
of doping, and have shown that the doping-dependent splitting of the
magnetization plateaux is a generic feature that goes beyond spin
$1/2$ Hubbard or $t-J$ models. We have tested the validity of the 
approach by verifying some of the well-known results for undoped
and doped spin-$1/2$ chains and ladders
\cite{plateaux_ladd,Plateaux_pmer,Cabra_irrational,Poilblanc2004,Roux_irrational,Roux_2007,Roux_PhD}.
Then, to illustrate the power of the approach, we have studied a doped
higher-spin system that would present a problem for a rigorous treatment 
via traditional bosonization technique. In contrast to the zero-magnetization 
case, initially treated by Shankar, here we have successfully used the path 
integral approach in the case of doped systems with nearest-neighbor 
hopping only. To the best of our knowledge, this work also offers the first 
description of a doped antiferromagnetic system in the presence of a magnetic 
field within the path integral approach.
Our analytical calculations have been supplemented by DMRG numerical 
work in two key examples of doped systems, the trimerized spin $1/2$ chain,  
and a spin $3/2$ chain.  All of our results are represented by the Eq.~(\ref{Main}), 
but in fact this approach is applicable far beyond the examples given in 
the paper. It appears that now we have a reasonably controlled 
technique to study the origin of magnetization plateaux in system 
with arbitrary spin, dimensionality and doping. This is to be contrasted 
to other techniques that are limited either to one dimension (such as 
bosonization or DMRG), or to spin-$1/2$ (such as numerical diagonalization 
in dimensions $D\geq 2$, or bond-operator/hard core bosons descriptions).

\acknowledgments

We would like to thank K. Totsuka for enlightening discussions and a special thanks to R.\ Ramazashvili 
for long and particularly helpful discussions and seminal contributions to this work. 
S. C. and P. P. would also like to thank A. Euverte with whom 
part of this work was initiated. Numerical simulations were performed using 
HPC resources from GENCI-IDRIS (Grant 2009-100225) and CALMIP.

%


\begin{thebibliography}{breitestes Label}
\bibitem{comment1} In fact, it can be shown that these two conditions are equivalent only in the case of a vortex-free configuration of the $\phi$ field.
\end{thebibliography}
\end{document}